\documentclass[english]{article}
\usepackage{ae,aecompl}

\usepackage[T1]{fontenc}
\usepackage[latin9]{inputenc}
\usepackage[letterpaper]{geometry}
\usepackage{color}
\usepackage{babel}

\usepackage{amsthm}
\usepackage{amsmath}
\usepackage{graphicx}
\usepackage{setspace}
\usepackage{amssymb}
\usepackage{esint}
\usepackage[unicode=true, pdfusetitle,
 bookmarks=false,
 breaklinks=false,pdfborder={0 0 1},backref=false,colorlinks=false]
 {hyperref}

\makeatletter

\providecommand{\tabularnewline}{\\}

  \newlength{\myFootnoteLabel}
  \renewcommand{\@makefntext}[1]{ %
  \noindent\makebox[\myFootnoteLabel][r]{\@makefnmark\ }%
  \parbox[t]{0.9\linewidth}{#1}%
  }

\makeatother

\begin{document}

\title{Lifshitz Solitons}

\author{R. B. Mann%
\thanks{email address: \texttt{\protect\href{mailto:rbmann@sciborg.uwaterloo.ca}{rbmann@sciborg.uwaterloo.ca}}%
}, L. Pegoraro%
\thanks{email address: \texttt{\protect\href{mailto:lpegorar@uwaterloo.ca}{lpegorar@uwaterloo.ca}}%
}, and M. Oltean%
\thanks{email address: \texttt{\protect\href{mailto:moltean@uwaterloo.ca}{moltean@uwaterloo.ca}}%
}\textit{\normalsize }\\
\textit{\normalsize Department of Physics and Astronomy, University
of Waterloo}\\
\textit{\normalsize 200 University Avenue West, Waterloo, Ontario,
Canada, N2L 3G1}}
\maketitle
\begin{abstract}
We numerically obtain a class of soliton solutions for Einstein gravity
in $(n+1)$ dimensions coupled to massive abelian gauge fields and
with a negative cosmological constant with Lifshitz asymptotic behaviour.
We find that for all $n\geq3$, a discrete set of magic values for
the charge density at the origin (guaranteeing an asymptotically Lifshitz
geometry) exists when the critical exponent associated with the Lifshitz
scaling is $z=2$; moreover, in all cases, a single magic value is
obtained for essentially every $1<z<2$, yet none when $z>2$ sufficiently.
\end{abstract}

\section{Introduction}

Since its proposal by Maldacena \cite{maldacena-1998}, the AdS/CFT
correspondence has proven to be of appreciably broad theoretical utility,
providing new lines of research into both quantum gravity and quantum
chromodynamics. It conjectures the existence of a holographic duality
between strongly interacting field theories and weakly coupled gravitational
dynamics in an asymptotically AdS bulk spacetime of one dimension
greater. 

This idea has been extended in recent years beyond high energy physics
to describe strongly coupled systems in condensed matter physics.
In particular, it has enjoyed useful applicability to theories that
model quantum critical behaviour \cite{hartnoll-2009,rokhsar-1988,ardonne-2004,vishwanath-2004}
characterized by Lifshitz scaling -- that is, a scaling transformation
of the form\begin{equation}
t\rightarrow\lambda^{z}t,\quad r\rightarrow\lambda^{-1}r,\quad\mathbf{x}\rightarrow\lambda\mathbf{x},\label{eq:1}\end{equation}
where $z\geq1$ is a dynamical critical exponent representing the
degree of anisotropy between space and time. For instance, when $z=2$,
the scaling symmetry given by (\ref{eq:1}) is associated with a $\left(2+1\right)$-dimensional
field theory of strongly correlated electron systems.

Such theories are conjectured \cite{kachru-2008} to be holographically
dual to gravitational theories whose solutions are asymptotic to the
so-called Lifshitz metric,\begin{equation}
ds^{2}=\ell^{2}\left(-r^{2z}dt^{2}+\frac{dr^{2}}{r^{2}}+r^{2}d\mathbf{x}^{2}\right),\label{eq:2}\end{equation}
where the coordinates $\left(t,r,x^{i}\right)$ are dimensionless
and the only length scale in the geometry is $\ell$. Metrics asymptotic
to (\ref{eq:2}) can be generated as solutions to the equations of
motion that follow from the action\begin{equation}
I=\frac{1}{16\pi}\int d^{n+1}x\sqrt{-g}\left(R-2\Lambda-\frac{1}{4}F_{\mu\nu}F^{\mu\nu}-\frac{1}{4}H_{\mu\nu}H^{\mu\nu}-\frac{C}{2}B_{\mu}B^{\mu}\right),\label{eq:3}\end{equation}
where $\Lambda$ is the cosmological constant, $F_{\mu\nu}=\partial_{[\mu}A_{\nu]}$
with $A_{\mu}$ representing the Maxwell gauge field, and $H_{\mu\nu}=\partial_{[\mu}B_{\nu]}$
is the field strength of the Proca field $B_{\mu}$ with mass $m^{2}=C$.

This dual theory, referred to as Lifshitz gravity, is known to describe
neutral and charged black holes \cite{danielsson-2009}\cite{mann:2009}\cite{Peet:2009}\cite{dehghani-2011}
whose metrics are asymptotic to the metric (\ref{eq:2}). It is possible
to replace the Proca field with higher-order curvature terms \cite{Dehghani:2010}\cite{Brenna:2011}\cite{Matulich-2011}
and attain black hole metrics with the same asymptotic structure.

Here we explore the possibility of obtaining soliton solutions to
the field equations that follow from the action (\ref{eq:3}) when
the Maxwell field is zero. Originally referred to as Lifshitz stars
\cite{danielsson-2009}, these objects have non-singular spacetime
geometries with the same asymptotics as their black hole counterparts.
We prefer to call them solitons since, unlike stars, there is no sharp
boundary between a vacuum and non-vacuum region. While black hole
solutions in $(n+1)$ dimensions \cite{dehghani-2011} and soliton
solutions in $(2+1)$ \cite{Gonzalez-2011} and
$(3+1)$ dimensions \cite{danielsson-2009} have already
been found, solitons in higher dimensions have not been investigated
thus far. 

 Here our central aim is to obtain soliton solutions
for general ($n,z$), and to discuss some of their consequences. For
any dimension $n$, we find that a single soliton solution to the
field equations exists for $z<2$, whereas for $z>2$ we are unable
to obtain any solutions. For $z=2$ we find a discrete set of soliton
solutions in any dimensionality. Our results are numerical, and so
we are able to obtain additional soliton solutions for $|z-2|<\epsilon$
for sufficiently small $\epsilon$ , where $\epsilon$ decreases with
increasing dimensionality. 

The rest of this paper is organized as follows. In Section 2, we give
the field equations for Lifshitz gravity and then reduce them to a
system of four first-order ODEs. In Section 3, we numerically solve
the equations of motion for $n=3,4,5,6$ and list the corresponding
magic values for the central charge density in each case. Section 4 concludes the
paper, and the Appendix gives the small-radius series coefficients
of the metric and gauge functions (needed to determine a set of initial
conditions for Section 3).

\section{The $(n+1)$-Dimensional Field Equations}

Given the action (\ref{eq:3}) with $A_{\mu}=0$, the variational
principle yields the field equations \cite{dehghani-2011}:\begin{align}
G_{\mu\nu}+\Lambda g_{\mu\nu} & =8\pi T_{\mu\nu},\label{eq:4}\\
\nabla^{\mu}H_{\mu\nu} & =CB_{\mu},\label{eq:5}\\
\partial_{[\mu}B_{\nu]} & =H_{\mu\nu},\label{eq:6}\end{align}
where the energy-momentum tensor of the gauge fields is\begin{equation}
T_{\mu\nu}=-\frac{1}{2}\left[\frac{1}{4}H_{\rho\sigma}H^{\rho\sigma}g_{\mu\nu}-H_{\,\mu}^{\rho}H_{\rho\nu}+C\left(\frac{1}{4}B_{\rho}B^{\rho}g_{\mu\nu}-B_{\mu}B_{\nu}\right)\right].\label{eq:8}\end{equation}
The general $(n+1)$-dimensional metric preserving the basic symmetries
(\ref{eq:1}) can be written as\begin{equation}
ds^{2}=\ell^{2}\left(-r^{2z}f^{2}\left(r\right)dt^{2}+\frac{g^{2}\left(r\right)dr^{2}}{r^{2}}+r^{2}d\Omega_{k}^{2}\right),\label{eq:9}\end{equation}
where \begin{equation}
d\Omega_{k}^{2}=\begin{cases}
{\displaystyle d\theta_{1}^{2}+\sum_{i=2}^{n-1}\prod_{j=1}^{i-1}\sin^{2}\theta_{j}d\theta_{i}^{2},} & k=1,\\
{\displaystyle d\theta_{1}^{2}+\sinh^{2}\theta_{1}\left(d\theta_{2}^{2}+\sum_{i=3}^{n-1}\prod_{j=2}^{i-1}\sin^{2}\theta_{j}d\theta_{i}^{2}\right),} & k=-1,\\
{\displaystyle \sum_{i=1}^{n-1}d\theta_{i}^{2},} & k=0\end{cases}\label{eq:10}\end{equation}
is the metric of an $(n-1)$-dimensional hypersurface with constant
curvature $(n-1)(n-2)k$.

The Proca field is assumed to be\begin{equation}
B_{t}=q\ell r^{z}f\left(r\right)j\left(r\right),\quad H_{tr}=q\ell zr^{z-1}g\left(r\right)h\left(r\right)f\left(r\right),\label{eq:11}\end{equation}
with all other components either vanishing or given by antisymmetrization.
The asymptotic conditions $f\left(r\right)$, $g\left(r\right)$,
$h\left(r\right)$, $j\left(r\right)\rightarrow1$ as $r\rightarrow\infty$
(required to ensure that any solutions obtained are asymptotic to
(\ref{eq:2})) impose the following constraints: \begin{equation}
C=\frac{\left(n-1\right)z}{\ell^{2}},\quad q^{2}=\frac{2\left(z-1\right)}{z},\quad\Lambda=-\frac{\left(z-1\right)^{2}+n\left(z-2\right)+n^{2}}{2\ell^{2}}.\label{eq:12}\end{equation}
It can be shown \cite{dehghani-2011} that the above reduce the field
equations (\ref{eq:4})-(\ref{eq:6}) to a system of four first-order
ODEs, \begin{align}
r\frac{df}{dr}= & \frac{f}{4\left(n-1\right)r^{2}}\Big\{2\left[\left(n-1\right)\left(z-1\right)j^{2}-z\left(z-1\right)h^{2}+\left(z-1\right)^{2}+n\left(z-2\right)+n^{2}\right]r^{2}g^{2}\nonumber \\
 & +2\left(n-1\right)\left[\left(n-2\right)k\ell^{2}g^{2}-\left(n+2z-2\right)r^{2}\right]\Big\},\label{eq:13}\\
r\frac{dg}{dr}= & \frac{g}{4\left(n-1\right)r^{2}}\Big\{2\left[\left(n-1\right)\left(z-1\right)j^{2}+z\left(z-1\right)h^{2}-\left(z-1\right)^{2}-n\left(z-2\right)-n^{2}\right]r^{2}g^{2}\nonumber \\
 & -2\left(n-1\right)\left[\left(n-2\right)k\ell^{2}g^{2}-nr^{2}\right]\Big\},\label{eq:14}\\
r\frac{dj}{dr}= & \frac{-j}{4\left(n-1\right)r^{2}}\Big\{2\left[\left(n-1\right)\left(z-1\right)j^{2}-z\left(z-1\right)h^{2}+\left(z-1\right)^{2}+n\left(z-2\right)+n^{2}\right]r^{2}g^{2}\nonumber \\
 & +2\left(n-1\right)\left[\left(n-2\right)k\ell^{2}g^{2}-\left(n-2\right)r^{2}\right]\Big\}+zgh,\label{eq:15}\\
r\frac{dh}{dr}= & \left(n-1\right)\left(jg-h\right).\label{eq:16}\end{align}
which, in general, cannot be solved analytically.

\section{Numerical Solutions for $(n+1)$-Dimensional Solitons}

Series solutions for the field equations at large $r$ have been previously
obtained \cite{danielsson-2009}\cite{mann:2009}\cite{Dehghani:2010}. 

Before we can numerically obtain soliton solutions to the equations
of motion (\ref{eq:13})-(\ref{eq:16}), we require a set of initial
conditions i.e. the values of $f\left(\varepsilon\right)$, $g\left(\varepsilon\right)$,
$j\left(\varepsilon\right)$ and $h\left(\varepsilon\right)$ for
some $0<\varepsilon\ll1$. \textcolor{black}{For soliton solutions
we demand that the metric be regular and that the Proca field have
vanishing field strength at the origin.} To this end, consider the
following series expansions for small $r$:\begin{equation}
f\left(r\right)=\frac{1}{r^{z}}\sum_{p=0}^{\infty}f_{p}r^{2p},\quad g\left(r\right)=r\sum_{p=0}^{\infty}g_{p}r^{2p},\quad j\left(r\right)=\sum_{p=0}^{\infty}j_{p}r^{2p},\quad h\left(r\right)=r\sum_{p=1}^{\infty}h_{p}r^{2p}.\label{eq:17}\end{equation}
It is possible to write all of the coefficients $\left\{ f_{p},g_{p},j_{p},h_{p}\right\} _{p=0}^{\infty}$
simply in terms of $f_{0}$ and $j_{0}$ (which is the Proca charge
density at the center of the soliton). In the Appendix, we give the
full expressions for the first three terms in the series for each
function in (\ref{eq:17}). We find that only for $k=1$ are the series
solutions finite and real, so we shall henceforth set $k=1$.

We can now numerically solve the system (\ref{eq:13})-(\ref{eq:16})
using the shooting method, taking as initial conditions (\ref{eq:18})-(\ref{eq:21})
(given in the Appendix) truncated after two terms, evaluated at $\varepsilon=10^{-6}$.
Furthermore, we rescale all quantities in units of $\ell$, effectively
setting $\ell=1$.

As previously noted, (\ref{eq:18})-(\ref{eq:21}) are determined
solely by two parameters, $f_{0}$ and $j_{0}$. The value of the
former can be easily assigned. Since $f_{0}$ appears only as an overall
factor in (\ref{eq:18}), the system (\ref{eq:13})-(\ref{eq:16})
can be solved by setting $f_{0}=1$ as an initial condition. If $g\left(r\right)$,
$h\left(r\right)$, $j\left(r\right)\rightarrow1$ and $f\left(r\right)\rightarrow c\neq1$
as $r\rightarrow\infty$, we simply change the initial value of $f\left(r\right)$,to
$f_{0}=1/c$ and so obtain the desired asymptotic behaviour. 

The determination of the latter, however, is not quite so trivial.
We find in general that, as for the $(3+1)$-dimensional case \cite{danielsson-2009},
the conditions $f\left(r\right)$, $g\left(r\right)$, $h\left(r\right)$,
$j\left(r\right)\rightarrow1$ as $r\rightarrow\infty$ can be satisfied
only for certain discrete values of $j_{0}$, known as `magic' values
\cite{danielsson-2009}. These correspond to the intercepts of the
function $j_{0}\gamma_{0}$; here, $\gamma_{0}:=j\left(r_{L}\right)-1$,
where $j\left(r\right)$ is the numerical solution to (\ref{eq:15})
dependent upon our choice of $j_{0}$, and $r_{L}$ is picked to be
very large. We furthermore expect these intercepts to coincide with
those of the functions%
\footnote{Although these intercepts are, of course, the same as just those of
$\alpha_{0}$, $\beta_{0}$ and $\gamma_{0}$, multiplication by $j_{0}$
makes our plots more readable. %
} $j_{0}\alpha_{0}$ and $j_{0}\beta_{0}$; here, $\alpha_{0}:=g\left(r_{L}\right)-1$
and $\beta_{0}:=h\left(r_{L}\right)-1$, where $g\left(r\right)$
and $h\left(r\right)$ are the numerical solutions to (\ref{eq:14})
and (\ref{eq:16}) respectively, again depending upon the value of
$j_{0}$. For $z=2$ in $(3+1)$-dimensions these intercepts correspond
to the removal of a zero mode in the large-$r$ linearized field equations. 

In all of our work, we use $r_{L}=10^{5}$ unless otherwise stated.
Thus, plotting $j_{0}\gamma_{0}$ (as a function of $j_{0}$) for
different $n$, and different $z$ for each $n$, will suffice to
give us the magic values, if any exist.

We will consider three cases: $1<z<2$, $z=2$ and $z>2$.

\subsection{Case I: $1<z<2$}

We find that for all $1<z<2$ and any $n\geq3$, a single magic value
of $j_{0}$ exists. Concordantly, here we essentially always observe
a single intercept of the function $j_{0}\gamma_{0}$. The only possible
exception is for $|z-2|$\textit{ small} (generally by less than $0.1$)
in which case we found numerically that more than one magic value
may exist. Such scenarios are qualitatively comparable to those where
$z$ is exactly $2$. Since we discuss this case at length in the
next subsection we shall not elaborate on the small $|z-2|$\textit{
}cases any further here. 

Figure 1 shows this for $n=3$ and $n=4$, with the situation being
qualitatively very similar for all other $n$. Moreover, in Figure
2 we give the magic values that we have computed for $n=3,4,5,6$
as a function of $1<z<2$. We see that (in any dimension) the magic
value increases with increasing $z$.

\noindent \begin{center}
\includegraphics[scale=0.38]{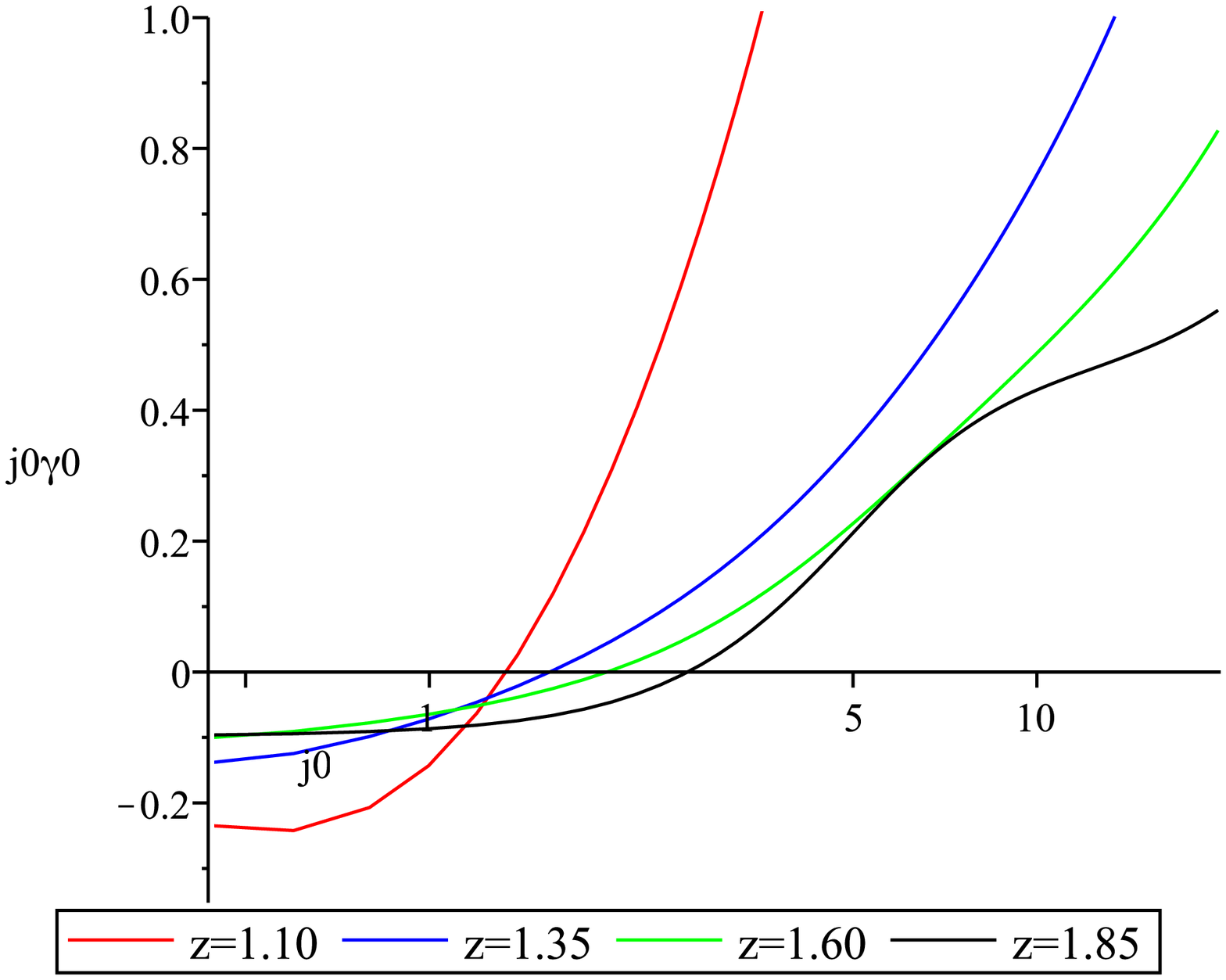}\quad{}\quad{}\quad{}\includegraphics[scale=0.38]{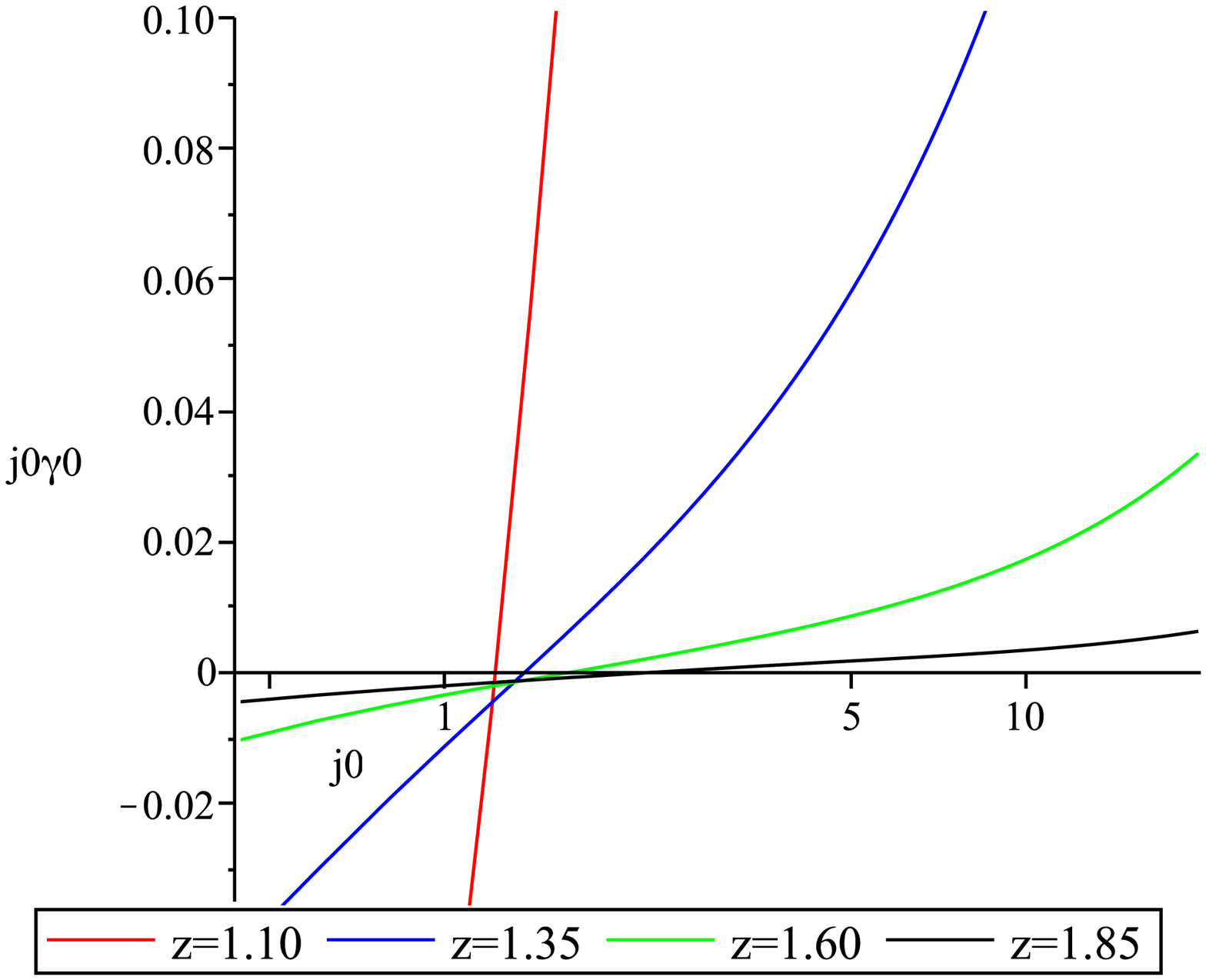}
\par\end{center}

\noindent \textbf{Figure 1:} Plots of $j_{0}\gamma_{0}$ as a function
of $j_{0}$ for different values of $1<z<2$. (The intercepts are
the magic values corresponding to each case.) Left: $n=3$. Right:
$n=4$. 

\noindent \begin{center}
\includegraphics[scale=0.48]{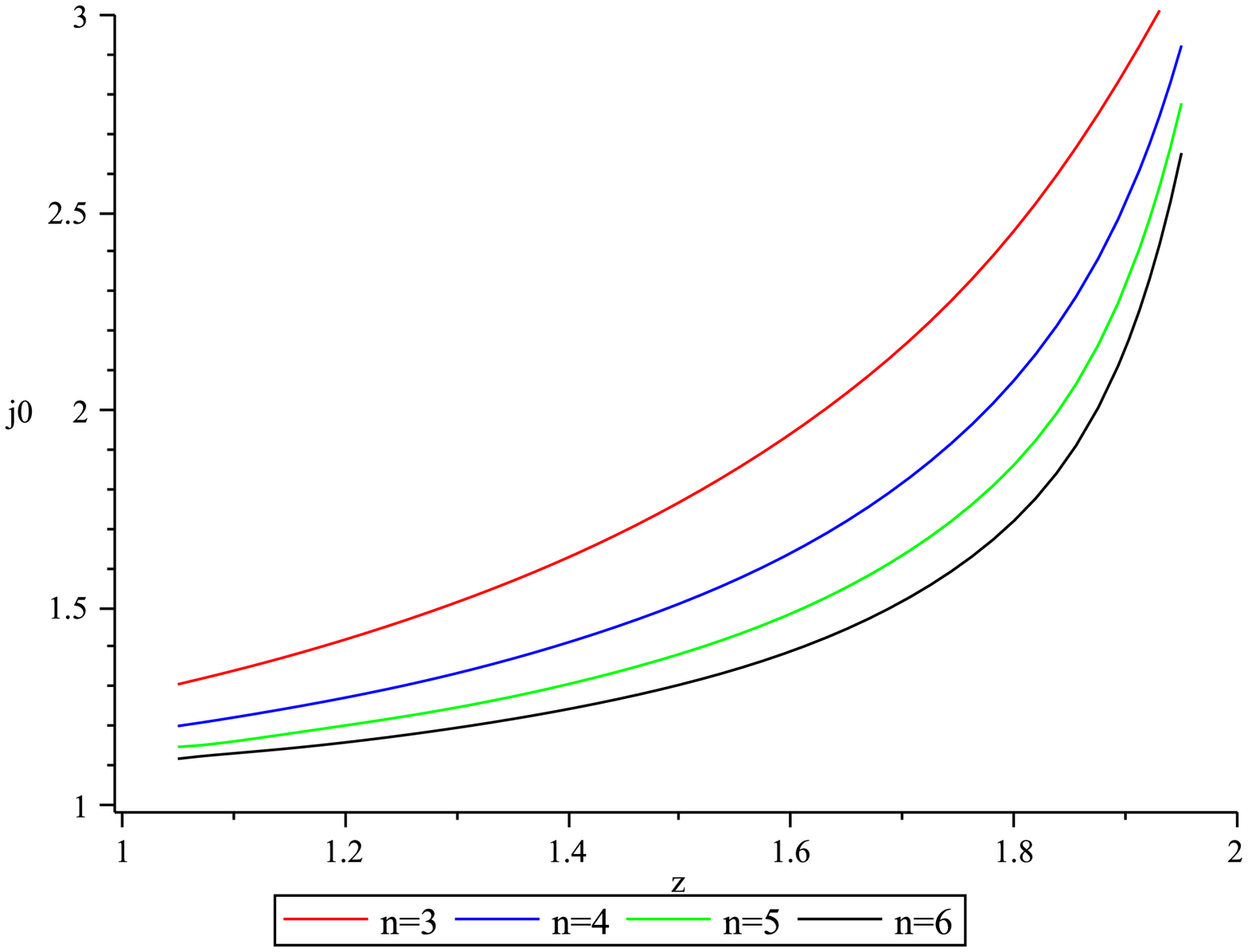}
\par\end{center}

\noindent \textbf{Figure 2:} Magic values of $j_{0}$ as a function
of $1<z<2$ for $n=3,4,5,6$. \textcolor{black}{This plot was obtained
by choosing, for each $n$, a discrete set of nine values of $z$
equally spaced along the given interval, numerically computing the
magic values for each (i.e. the intercepts corresponding to the kinds
of graphs depicted in Figure 1), and then polynomially interpolating
between them.}

\subsection{Case II: $z=2$}

In this case, a set of magic values exist for any $n$ for all cases
we numerically investigated. \textcolor{black}{Figure 3a depicts the
functions $j_{0}\alpha_{0}$, $j_{0}\beta_{0}$ and $j_{0}\gamma_{0}$
for $n=4,5$ while $j_{0}\gamma_{0}$ for $n=6$ is plotted Figure
3b. We find that the larger the value of $n$, the more rapidly oscillating
(and hence numerically unstable) these functions become -- and also,
the more difficult we find it to obtain convergence of their intercepts
with satisfactory accuracy. }

\textcolor{black}{Figure 4 lists the numerical values thereof for
$n=3,4,5,6$, i.e. the magic values in each dimension. For the plot
of $j_{0}\gamma_{0}$ }corresponding to $n=3$, see \cite{danielsson-2009};
note that in this case, we recover the same magic values as those
obtained therein.

Furthermore, we plot the metric and gauge functions corresponding
to the lowest magic value for $n=4,5,6$ in Figure 5a. Accordingly,
all of these are observed to converge to $1$ as $r\rightarrow\infty$
and, as generally expected for solitons, we see that $g\left(r\right)$
and $h\left(r\right)$ vanish as $r\rightarrow0$. In addition, the
function $r^{z}f\left(r\right)$, again for different $n$, is plotted
in Figure 5b.

\noindent \begin{center}
\includegraphics[scale=0.38]{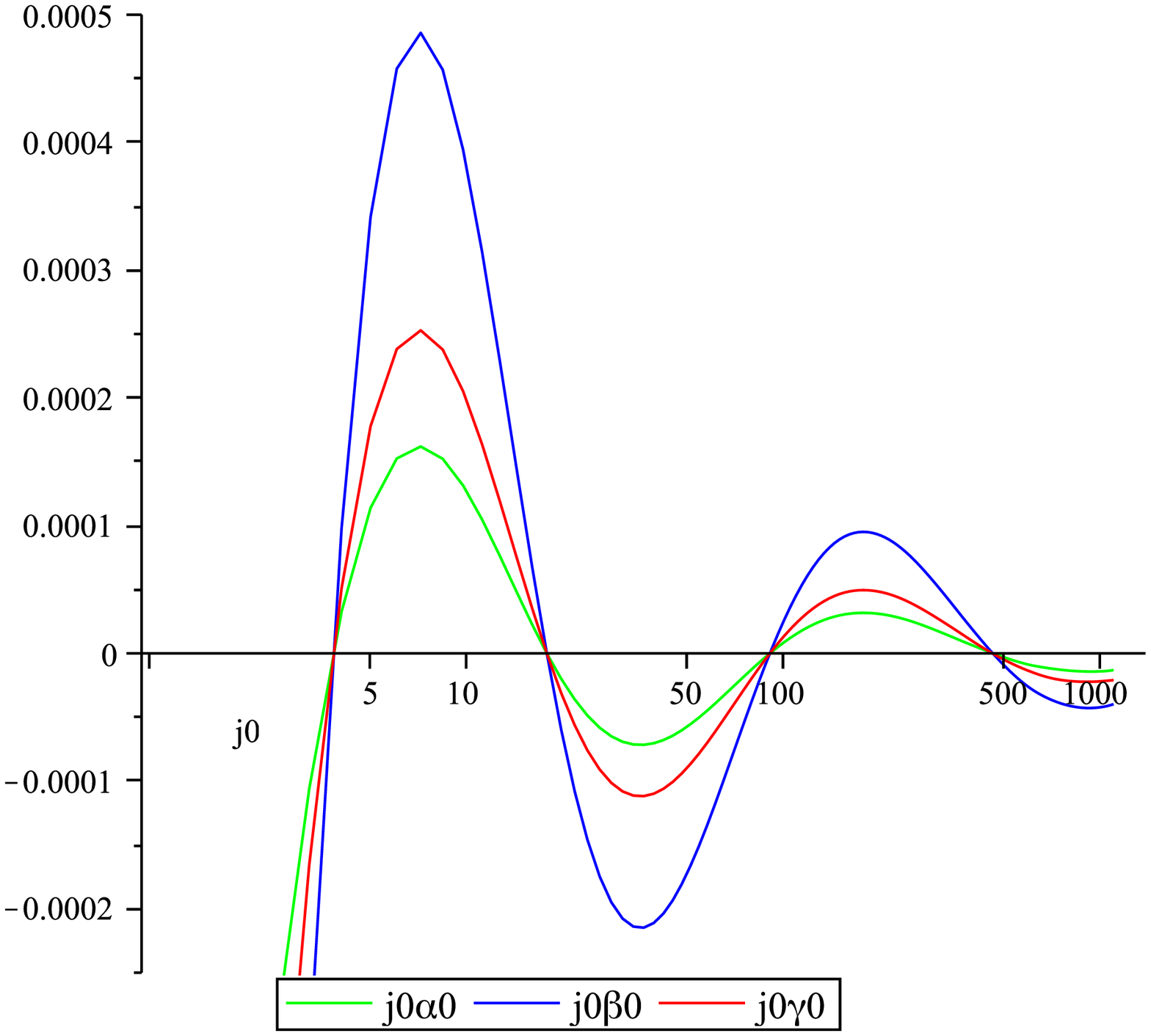}\quad{}\quad{}\quad{}\includegraphics[scale=0.38]{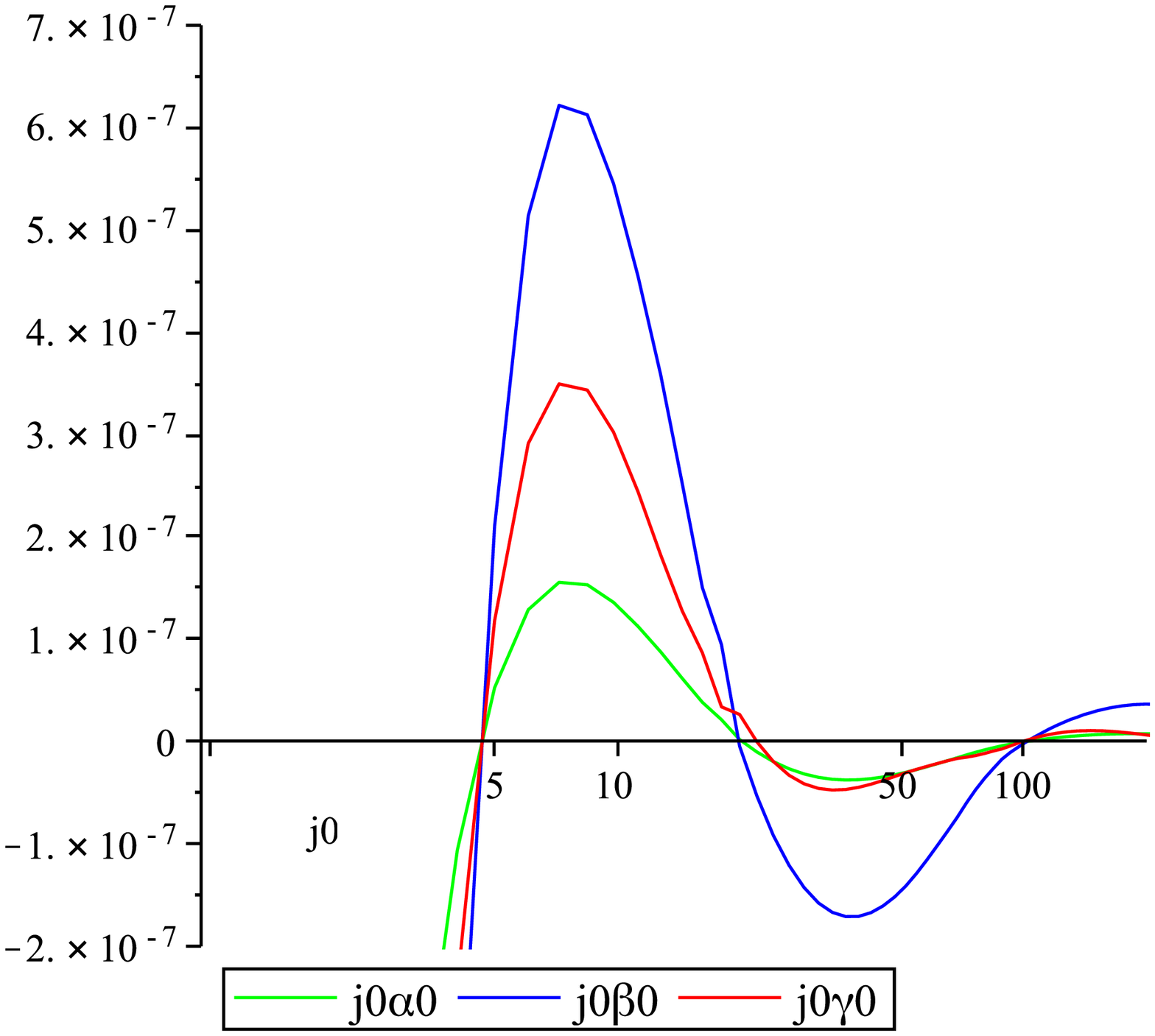}
\par\end{center}

\noindent \textbf{Figure 3a:} Plots of $j_{0}\alpha_{0}$, $j_{0}\beta_{0}$
and $j_{0}\gamma_{0}$ as functions of $j_{0}$ for $z=2$. Left:
$n=4$. As expected, all three functions are observed to have the
same intercepts. Right: $n=5$ with $r_{L}=10^{6}$. In this case,
convergence of the three functions to the same intercepts cannot be
obtained numerically quite as accurately. In particular, the second
intercept (i.e. magic value) is $20.3$ for $j_{0}\alpha_{0}$ and
$j_{0}\beta_{0}$, but $22.0$ for $j_{0}\gamma_{0}$; the rest, however,
are found to be the same.

\noindent \begin{center}
\includegraphics[scale=0.45]{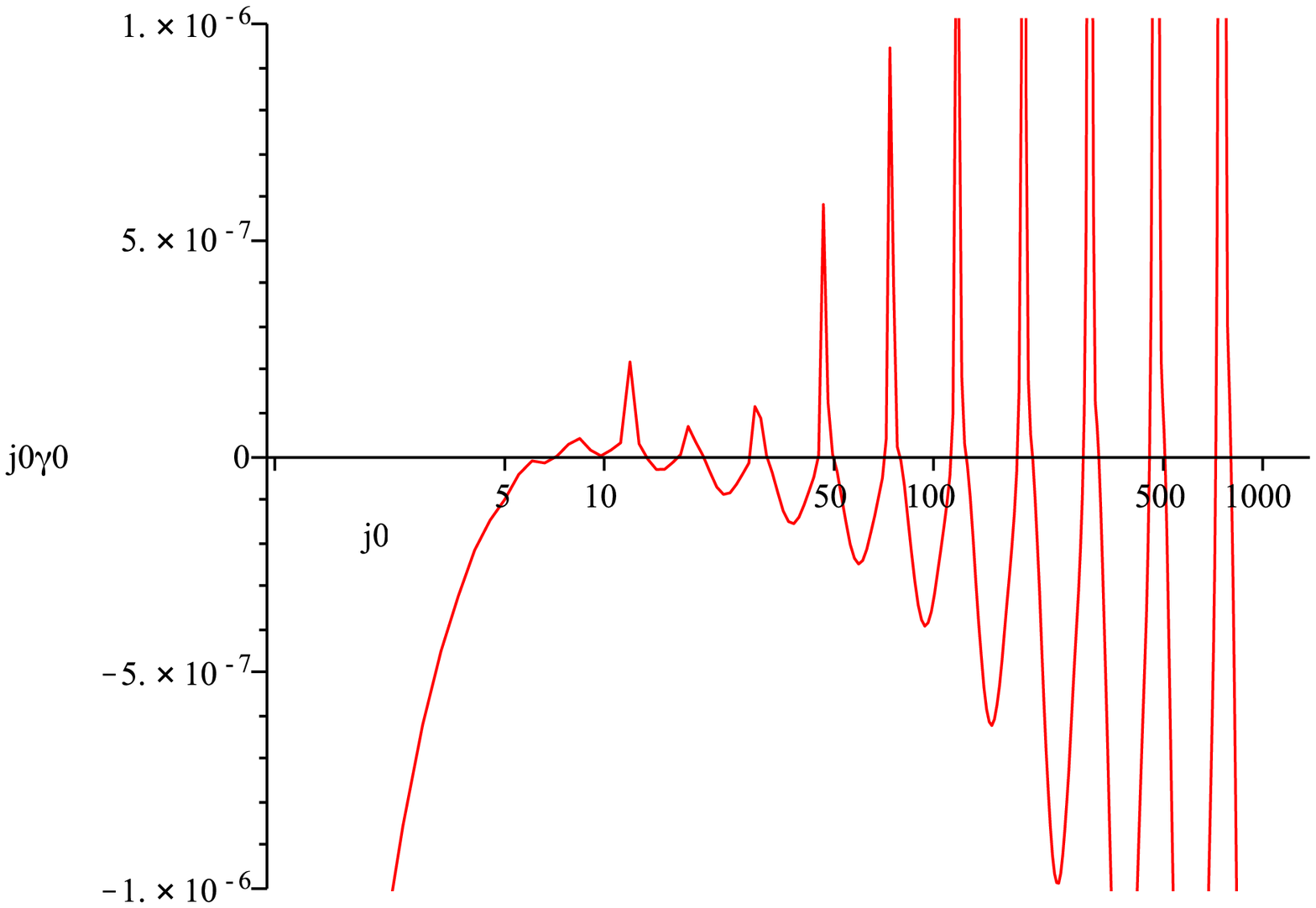}
\par\end{center}

\noindent \textbf{Figure 3b:} \textcolor{black}{Plot of $j_{0}\gamma_{0}$
when $n=6$. Numerical convergence of the intercepts of the above
plot with those of $j_{0}\alpha_{0}$ and $j_{0}\beta_{0}$ is not
very accurate in this case (becoming even worse for $n>6$), and so
here we base the list of magic values given in Figure 4 solely on
the former.}

\begin{spacing}{1.5}
\noindent \begin{center}
\begin{tabular}{|c|c|}
\hline 
$n$ & Magic values for $j_{0}$ when $z=2$\tabularnewline
\hline 
3 & $3.59,\;21.8,\;1.34\times10^{2},\;...$\tabularnewline
\hline 
4 & $3.80,\;18.0,\;91.2,\;...$\tabularnewline
\hline 
5 & $4.60,\;20.3,\;1.01\times10^{2},\;...$\tabularnewline
\hline 
6 & $7.20,\;9.74,\;13.5,\;...$\tabularnewline
\hline
\end{tabular}
\par\end{center}
\end{spacing}

\noindent \textbf{Figure 4:} The three lowest magic values for $z=2$
and $n=3,4,5,6$.

\noindent \begin{center}
\includegraphics[scale=0.35]{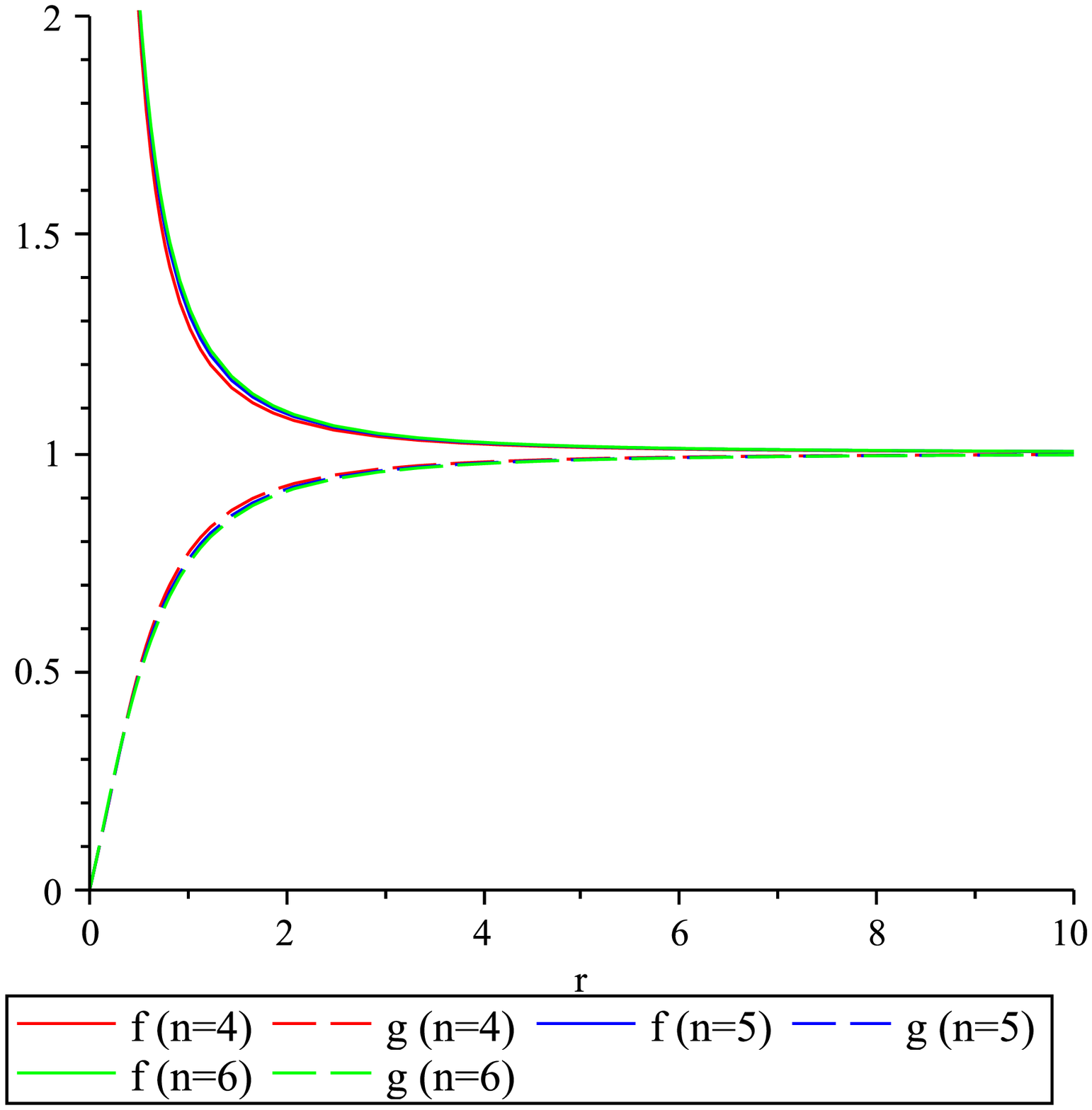}\quad{}\quad{}\quad{}\includegraphics[scale=0.35]{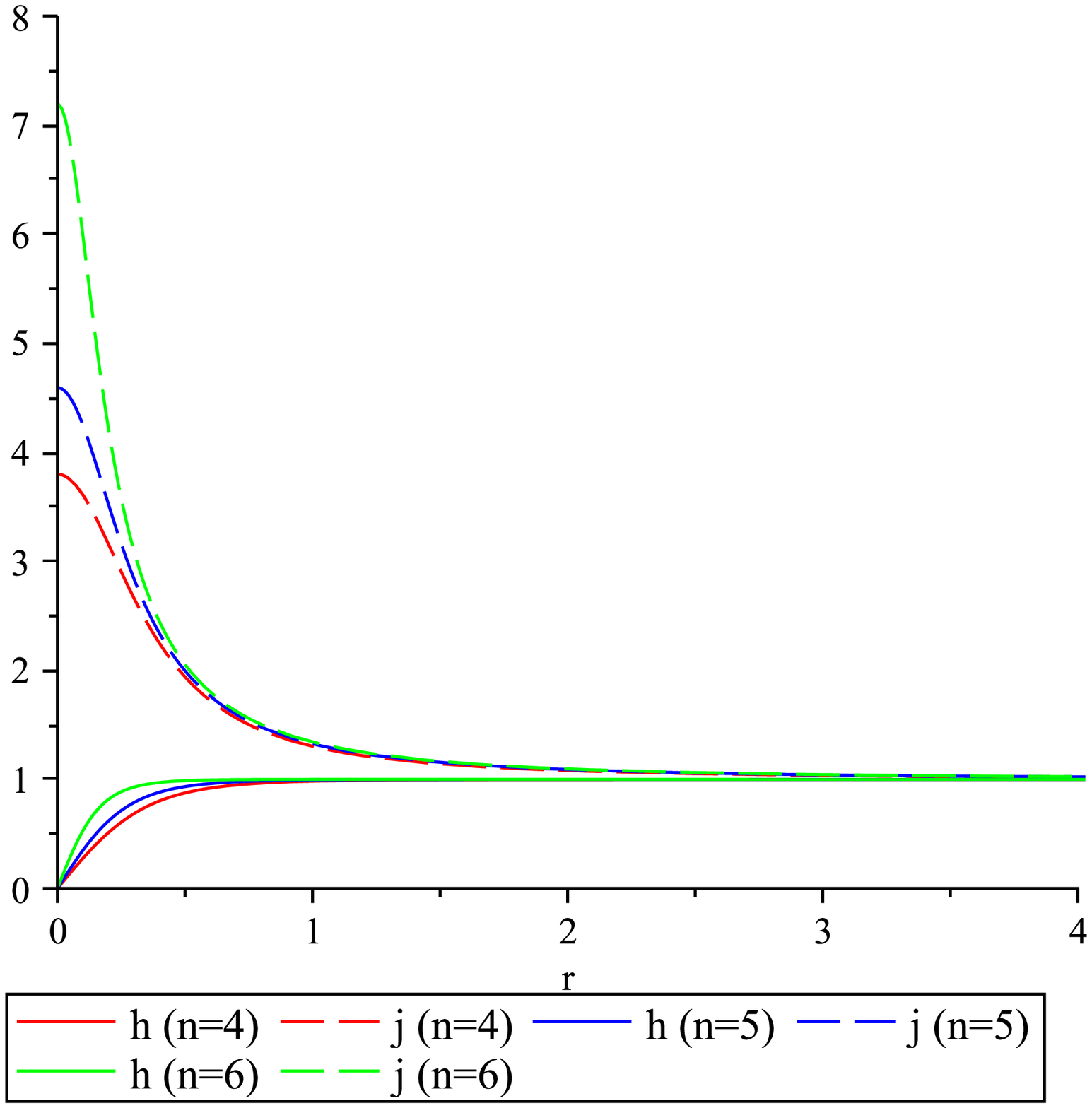}
\par\end{center}

\noindent \textbf{Figure 5a:} The metric and gauge functions corresponding
to the lowest magic value for $n=4,5,6$. Left: The metric functions
$f\left(r\right)$ and $g\left(r\right)$. Right: The gauge functions
$h\left(r\right)$ and $j\left(r\right)$. 

\noindent \begin{center}
\includegraphics[scale=0.4]{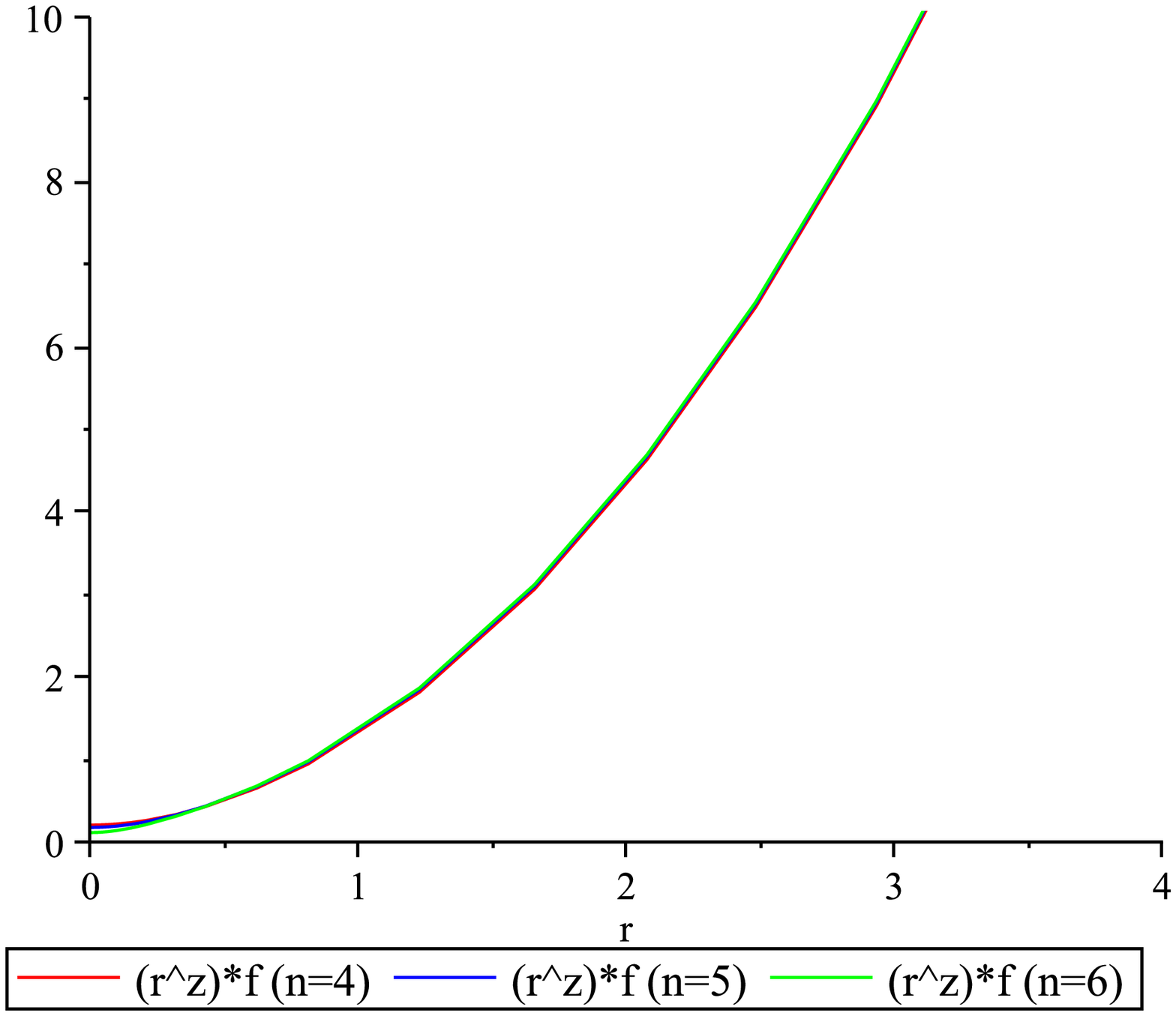}
\par\end{center}

\noindent \textbf{Figure 5b:} The function $r^{z}f\left(r\right)$
for $n=4,5,6$, illustrating that we indeed have soliton solutions.

\subsection{Case III: $z>2$}

While (multiple) intercepts of the function $j_{0}\gamma_{0}$ can
still be obtained if $z$ is made only\textit{ very slightly} greater
than $2$, none are seen to occur, in any dimension, once $z$ becomes
appreciably greater. In other words, we found no more magic values
for $z>2$. In particular, none exist for the zero modes (i.e. for
$z=n-1$) of any $n>3$. In Figures 6a, 6b and 6c we plot $j_{0}\gamma_{0}$
for, respectively, $n=4,5,6$ and various $z>2$. We find that magic
values can still be obtained (i.e. intercepts of these graphs exist)
only if, for instance, we have approximately $z<2.021$ when $n=4$,
$z<2.005$ when $n=5$, and $z<2.002$ when $n=6$. \textcolor{black}{The
analogous plots of $j_{0}\alpha_{0}$ and $j_{0}\beta_{0}$ are found
to be qualitatively similar, and so we have not included them here.}

\noindent \begin{center}
\includegraphics[scale=0.38]{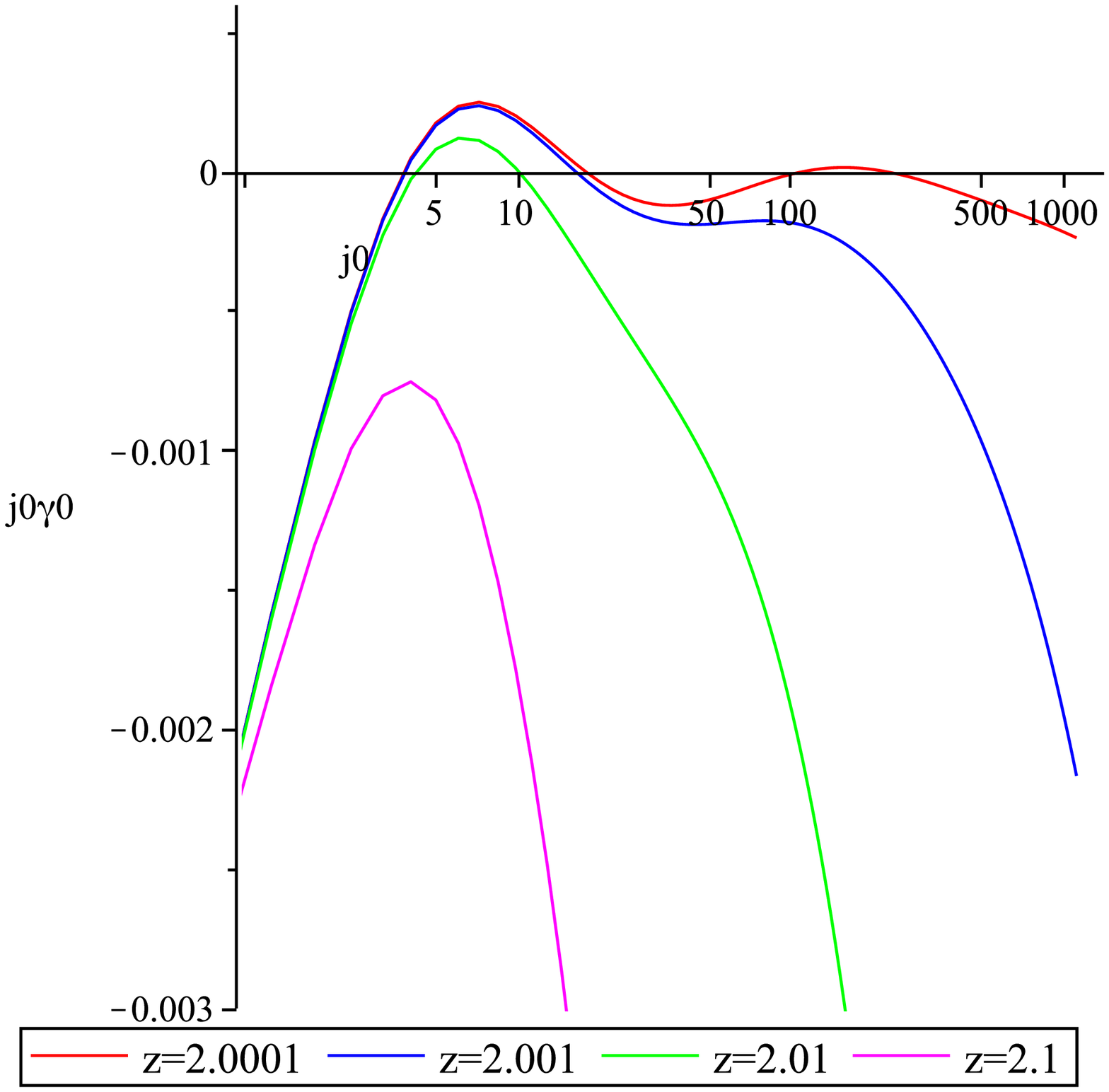}\quad{}\quad{}\quad{}\includegraphics[scale=0.38]{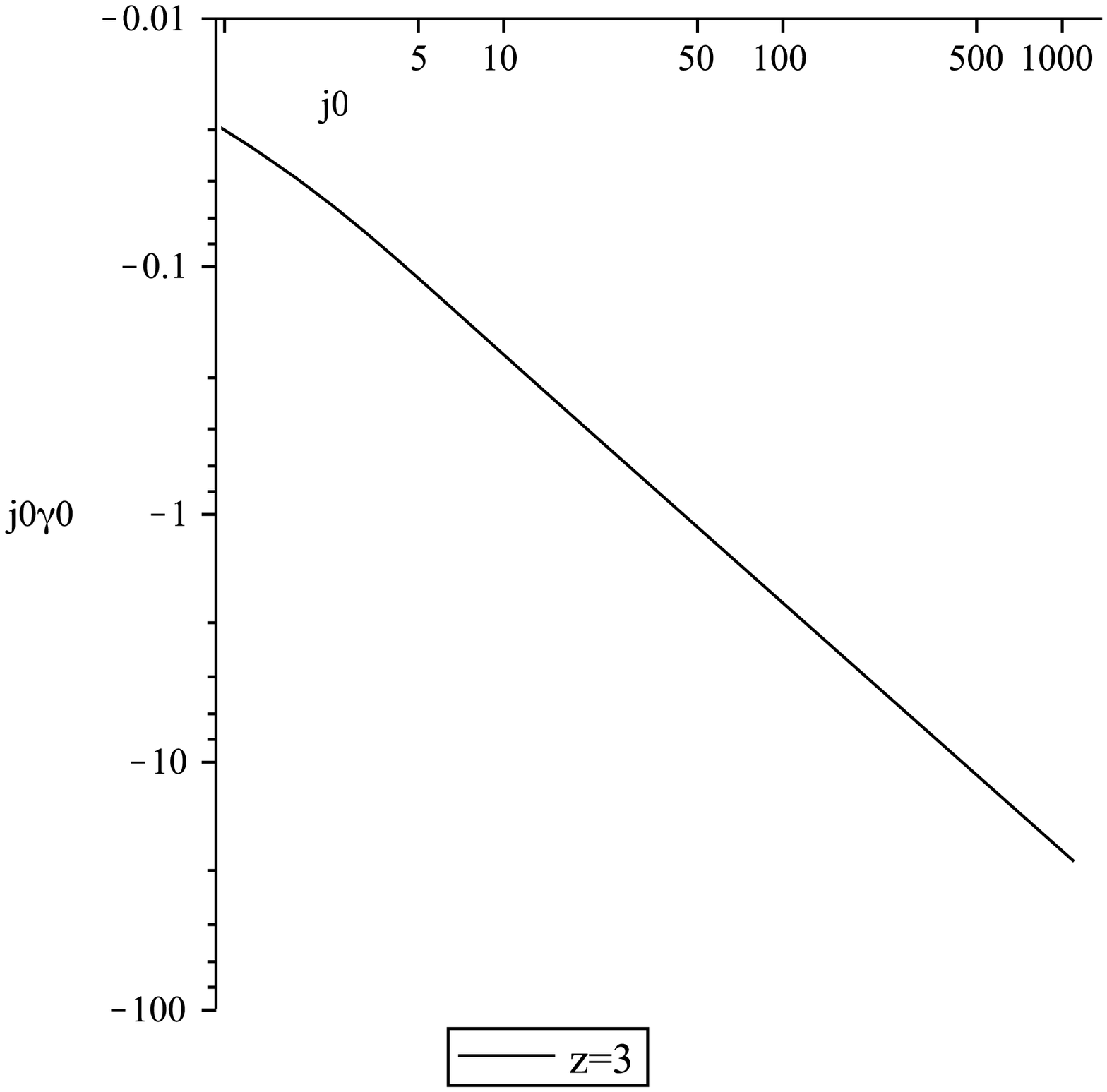}
\par\end{center}

\noindent \textbf{Figure 6a:} Plots of $j_{0}\gamma_{0}$ as a function
of $j_{0}$ for $n=4$ and various $z>2$.

\noindent \begin{center}
\includegraphics[scale=0.38]{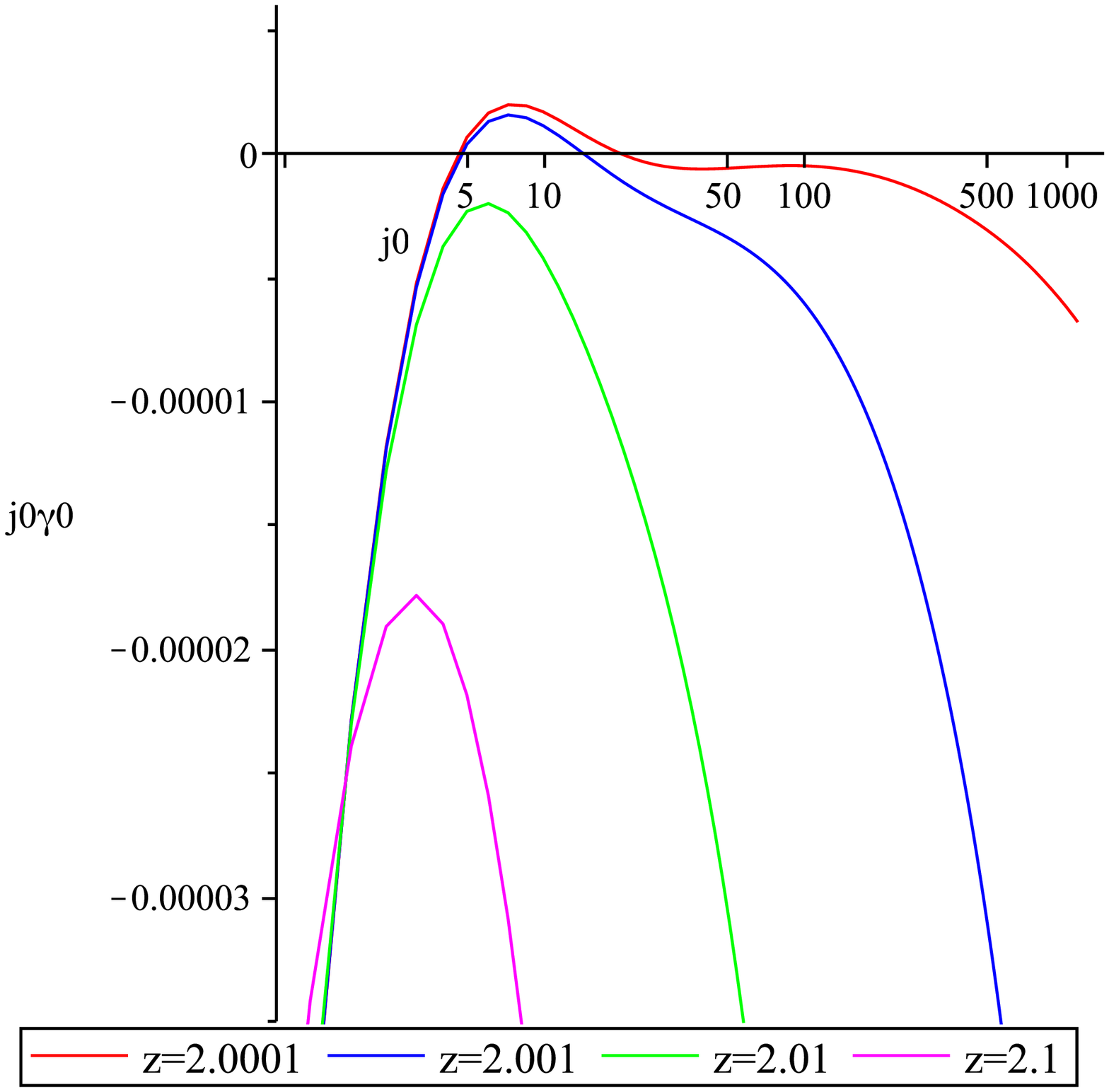}\quad{}\quad{}\quad{}\includegraphics[scale=0.38]{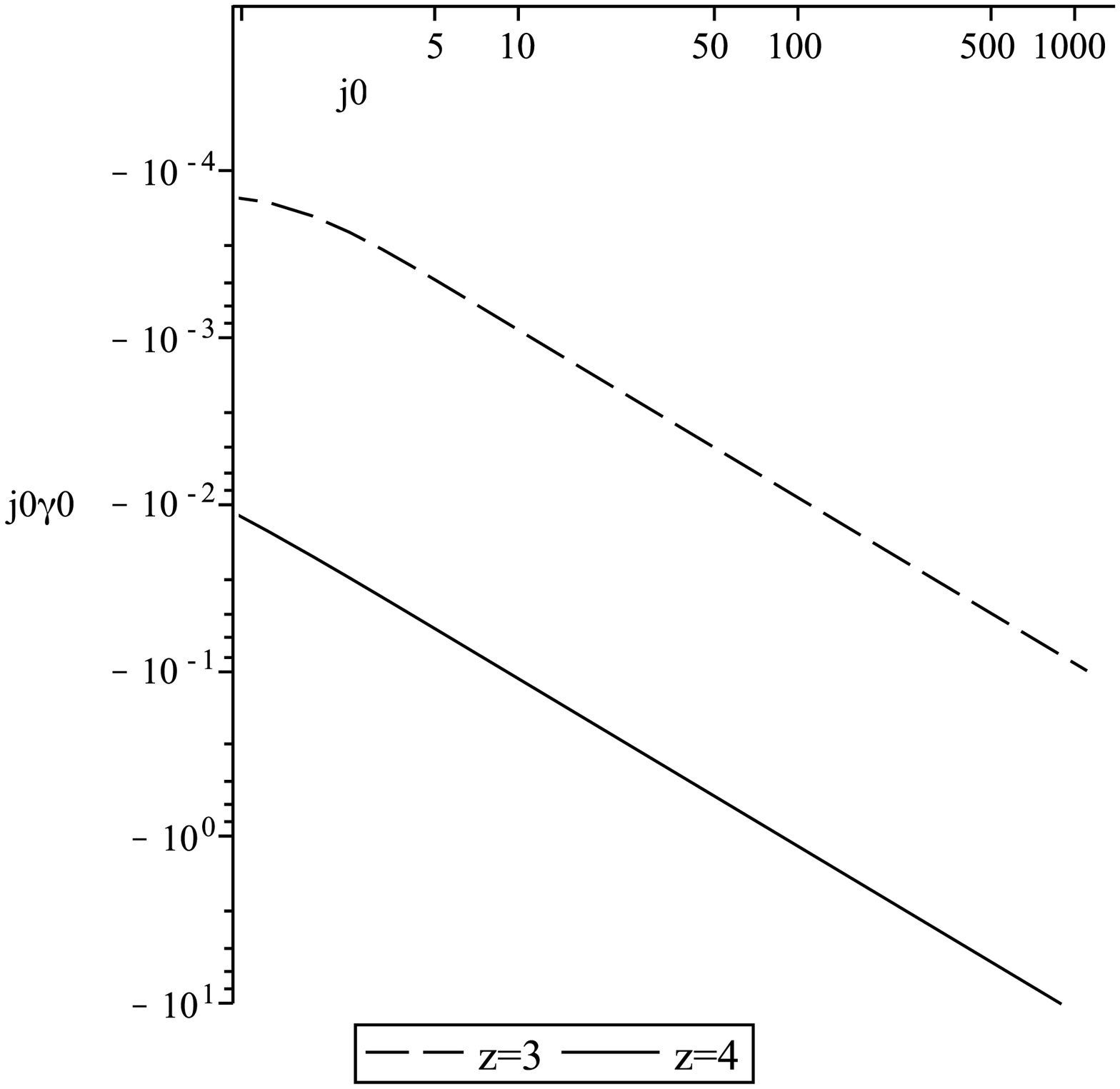}
\par\end{center}

\noindent \textbf{Figure 6b:} Plots of $j_{0}\gamma_{0}$ as a function
of $j_{0}$ for $n=5$ and various $z>2$.

\noindent \begin{center}
\includegraphics[scale=0.38]{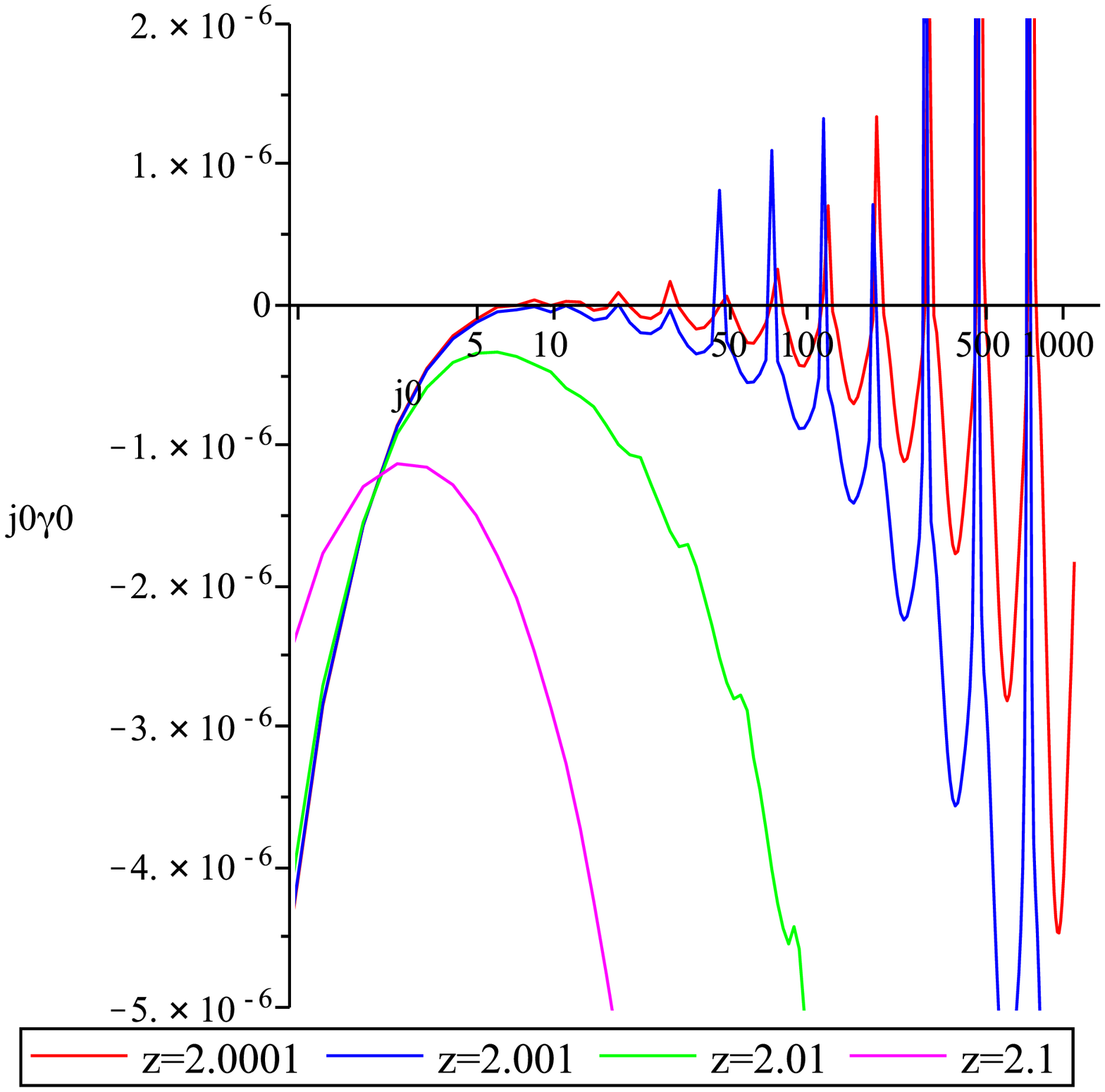}\quad{}\quad{}\quad{}\includegraphics[scale=0.38]{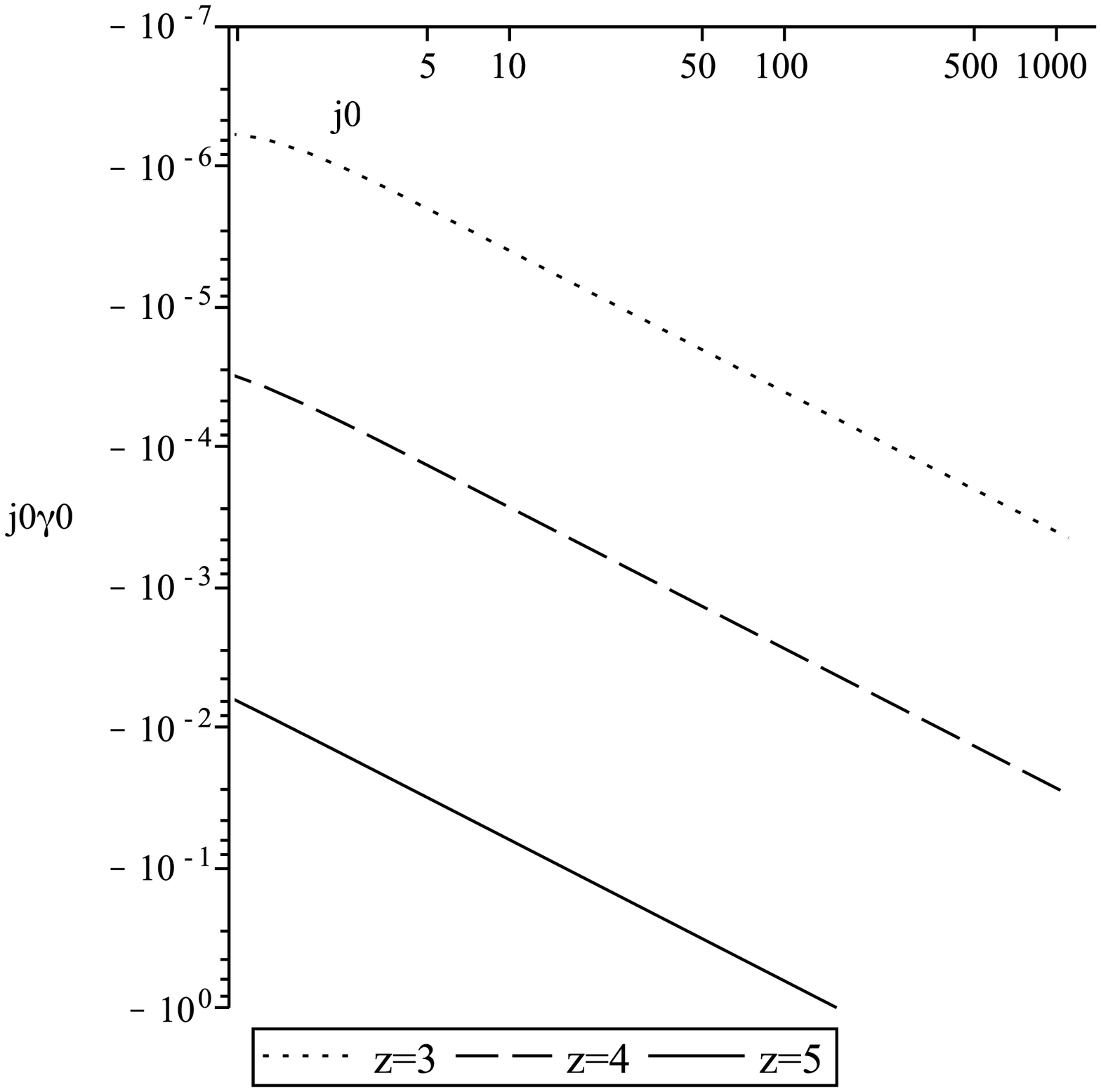}
\par\end{center}

\noindent \textbf{Figure 6c:} Plots of $j_{0}\gamma_{0}$ as a function
of $j_{0}$ for $n=6$ and various $z>2$.

\section{Conclusion}

We have searched for soliton solutions in asymptotically Lifshitz
spacetimes from $\left(3+1\right)$ \cite{danielsson-2009} to $\left(n+1\right)$
dimensions. We have found that such solutions do exist, but with a
somewhat surprising consequence: namely, solutions exist \textit{only}
when the critical exponent associated with the Lifshitz scaling is
(very close to) $2$, or smaller. In particular, $1<z<2$ and $z\approx2$
(to within at most $0.021$, $\forall n\geq4$) yield, respectively,
a single magic value and a discrete set of magic values, in any dimension.
But, once $z$ exceeds $2$ sufficiently, no more magic values --
and hence no more solutions for the metric and gauge functions --
can be numerically found. This means, therefore, that no soliton solutions
exist for the zero modes of any $n>3$. 

It would be interesting to understand in greater depth why $z=2$
is such a special point in parameter space in all of the dimensions
we investigated. Are such solutions stable, or will they undergo collapse
to a black hole? The relationship between these solutions and the
general (in)stability of asymptotically Lifshitz spacetimes \cite{Copsey}
would be another interesting subject to investigate.

\section*{Acknowledgements}

M. O. acknowledges support from the University of Waterloo Faculty
of Science and Faculty of Mathematics. This work was supported in
part by the Natural Sciences and Engineering Research Council of Canada.

\section*{Appendix}

It can easily be verified that\begin{eqnarray}
f\left(r\right) & = & \frac{1}{r^{z}}\left(f_{0}+f_{1}r^{2}+f_{2}r^{4}\right),\label{eq:18}\\
g\left(r\right) & = & r\left(g_{0}+g_{1}r^{2}+g_{2}r^{4}\right),\label{eq:19}\\
j\left(r\right) & = & j_{0}+j_{1}r^{2}+j_{2}r^{4},\label{eq:20}\\
h\left(r\right) & = & r\left(h_{0}+h_{1}r^{2}+h_{2}r^{4}\right),\label{eq:21}\end{eqnarray}
where\begin{align*}
f_{1}= & \frac{f_{0}}{2k\ell^{2}n\left(n-1\right)}\left\{ \left[\left(n-1\right)^{2}+z\left(n-2\right)+z^{2}\right]+\left[\left(z-1\right)\left(n-1\right)^{2}\right]j_{0}^{2}\right\} ,\\
f_{2}= & \frac{-f_{0}}{8\ell^{4}k^{2}\left(n-1\right)^{2}n^{2}\left(n+2\right)}\bigg\{\Big[2\left(z-1\right)^{4}+\left(z^{2}+2z-7\right)\left(z-1\right)^{2}n+2\left(z^{3}-z^{2}-3z+4\right)n^{2}\\
 & +\left(3z^{2}-4z-2\right)n^{3}+2\left(z-1\right)n^{4}+n^{5}\Big]+\left(z-1\right)\Big[4\left(z^{2}-3z+1\right)-2\left(z^{2}-10z+5\right)n-4\left(z^{2}+z-1\right)n^{2}\\
 & +2\left(z-1\right)\left(z-4\right)n^{3}+8\left(z-1\right)n^{4}-2\left(z-1\right)n^{5}\Big]j_{0}^{2}+\left(z-1\right)^{2}\left[2-3n-4n^{2}+10n^{3}-6n^{4}+n^{5}\right]j_{0}^{4}\bigg\},\end{align*}
\begin{align*}
g_{0}= & \frac{1}{k^{1/2}\ell},\\
g_{1}= & \frac{-1}{2k^{3/2}\ell^{3}n\left(n-1\right)}\left\{ \left[n^{2}+1-\left(2-z\right)\left(n+z\right)\right]-\left[\left(z-1\right)\left(n-1\right)\right]j_{0}^{2}\right\} ,\\
g_{2}= & \frac{-1}{8\ell^{5}k^{5/2}\left(n-1\right)^{2}n^{2}\left(n+2\right)}\bigg\{\Big[-6\left(z-1\right)^{4}-3\left(z^{2}+2z-7\right)\left(z-1\right)^{2}n+\left(-6z^{3}+6z^{2}+18z-24\right)n^{2}\\
 & +\left(-9z^{2}+12z+6\right)n^{3}+6\left(-z+1\right)n^{4}-3n^{5}\Big]+2\left(z-1\right)\Big[-2\left(3z^{2}-7z+3\right)+\left(z^{2}-12z+13\right)n\\
 & +\left(5z^{2}-9z-3\right)n^{2}+9\left(z-1\right)n^{3}+\left(-2z+5\right)n^{4}\Big]j_{0}^{2}+\left(z-1\right)^{2}\left[\left(n-1\right)^{2}\left(4n^{2}-7n-6\right)\right]j_{0}^{4}\bigg\},\end{align*}
\begin{align*}
j_{1}= & \frac{-j_{0}}{2k\ell^{2}n\left(n-1\right)}\left\{ \left[\left(n-1\right)^{2}+z\left(3n-n^{2}+z-3\right)\right]+\left[\left(z-1\right)\left(n-1\right)^{2}\right]j_{0}^{2}\right\} ,\\
j_{2}= & \frac{j_{0}}{8\ell^{4}k^{2}\left(n-1\right)^{2}n^{2}\left(n+2\right)}\bigg\{\Big[2\left(3z^{2}-8z+3\right)\left(z-1\right)^{2}+\left(3z^{4}+4z^{3}-41z^{2}+60z-21\right)n\\
 & +\left(10z^{3}-14z^{2}-26z+24\right)n^{2}+\left(-4z^{3}+27z^{2}-20z-6\right)n^{3}+\left(-8z^{2}+18z-6\right)n^{4}+\left(z-1\right)\left(z-3\right)n^{5}\Big]\\
 & +2\left(z-1\right)\Big[2\left(3z^{2}-8z+3\right)-\left(7z^{2}-31z+19\right)n-2\left(z^{2}+5z-9\right)n^{2}+z\left(3z-13\right)n^{3}+2\left(5z-4\right)n^{4}\\
 & -2\left(2z-3\right)n^{5}\Big]j_{0}^{2}+\left(z-1\right)^{2}\left[\left(n-1\right)^{3}\left(3n^{2}-n-6\right)\right]j_{0}^{4}\bigg\},\end{align*}
\begin{align*}
h_{0}= & \frac{j_{0}\left(n-1\right)}{k^{1/2}n\ell},\\
h_{1}= & \frac{-j_{0}}{2k^{3/2}\ell^{3}n\left(n+2\right)}\left\{ \left[n^{2}\left(2-z\right)+4n\left(z-1\right)+2z^{2}-5z+2\right]+\left[\left(z-1\right)\left(n^{2}-3n+2\right)\right]j_{0}^{2}\right\} ,\\
h_{2}= & \frac{j_{0}}{8\ell^{5}k^{5/2}\left(n-1\right)\left(n+4\right)n^{2}\left(n+2\right)}\bigg\{\Big[8\left(2z-1\right)\left(z-2\right)\left(z-1\right)^{2}+\left(8z^{4}+10z^{3}-107z^{2}+154z-56\right)n\\
 & +\left(20z^{3}-18z^{2}-72z+64\right)n^{2}+\left(-6z^{3}+46z^{2}-36z-16\right)n^{3}+\left(-10z^{2}+32z-16\right)n^{4}+\left(z-2\right)\left(z-4\right)n^{5}\Big]\\
 & +4\left(z-1\right)\Big[4(2z-1)(z-2)-2\left(3z^{2}-15z+11\right)n-\left(4z^{2}+3z-16\right)n^{2}+\left(2z^{2}-13z+4\right)n^{3}+\left(7z-8\right)n^{4}\\
 & +\left(-z+2\right)n^{5}\Big]j_{0}^{2}+\left(z-1\right)^{2}\left[\left(n-1\right)^{2}\left(n-2\right)\left(3n^{2}-4n-8\right)\right]j_{0}^{4}\bigg\},\end{align*}
satisfy the four ODEs (\ref{eq:13})-(\ref{eq:16}), at least up to
sixth order in $r$. In Figure 7, we compare the above series solutions
with the numerical results obtained (in Figure 5a) for the lowest
magic value when $n=4$, and find that they are in good agreement
up to at least $r\thickapprox0.1$. These plots are found to be qualitatively
similar for all other $n>4$, and hence we refrain from showing them
as well.

\noindent \begin{center}
\includegraphics[scale=0.38]{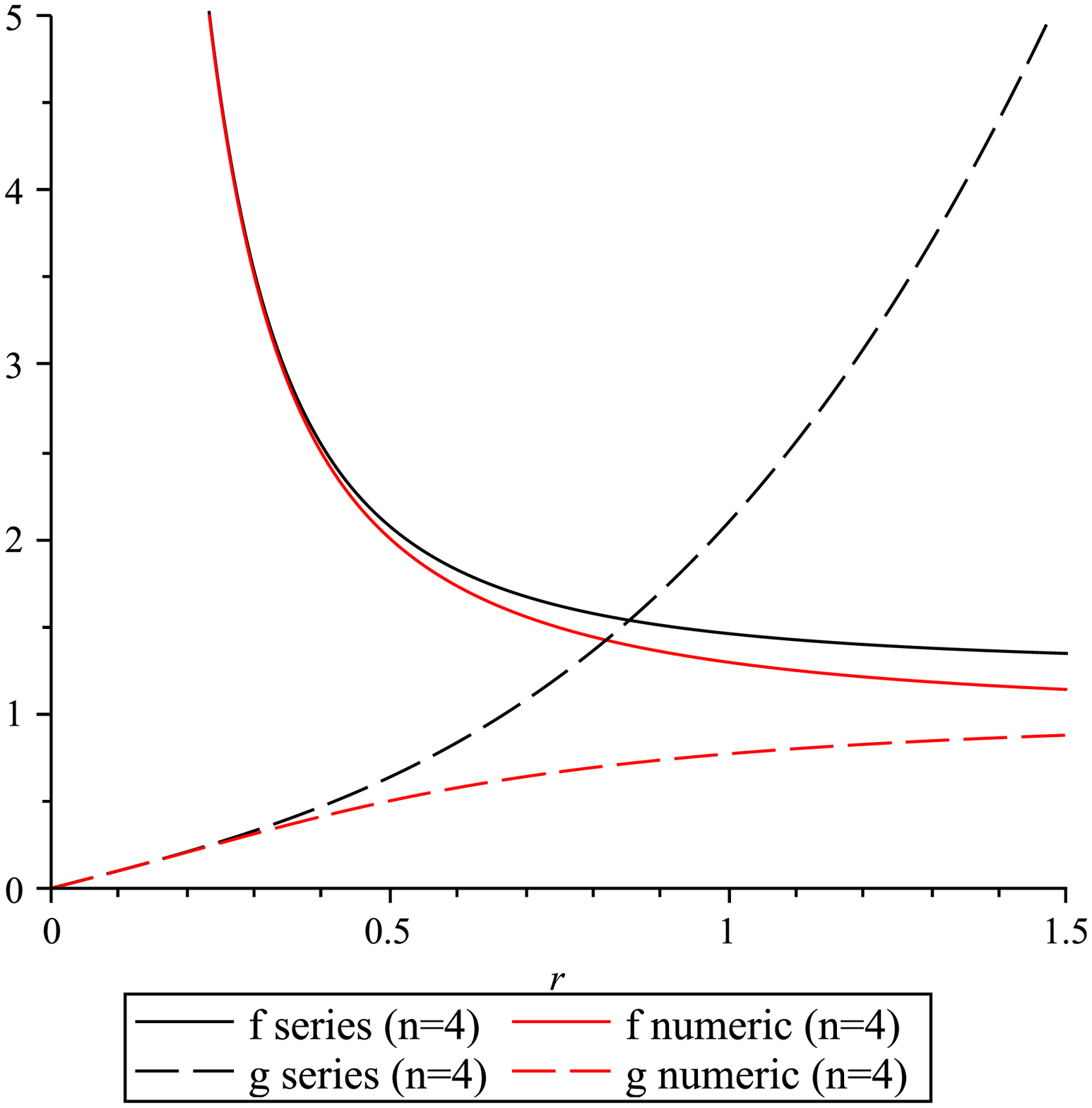}\quad{}\quad{}\quad{}\includegraphics[scale=0.38]{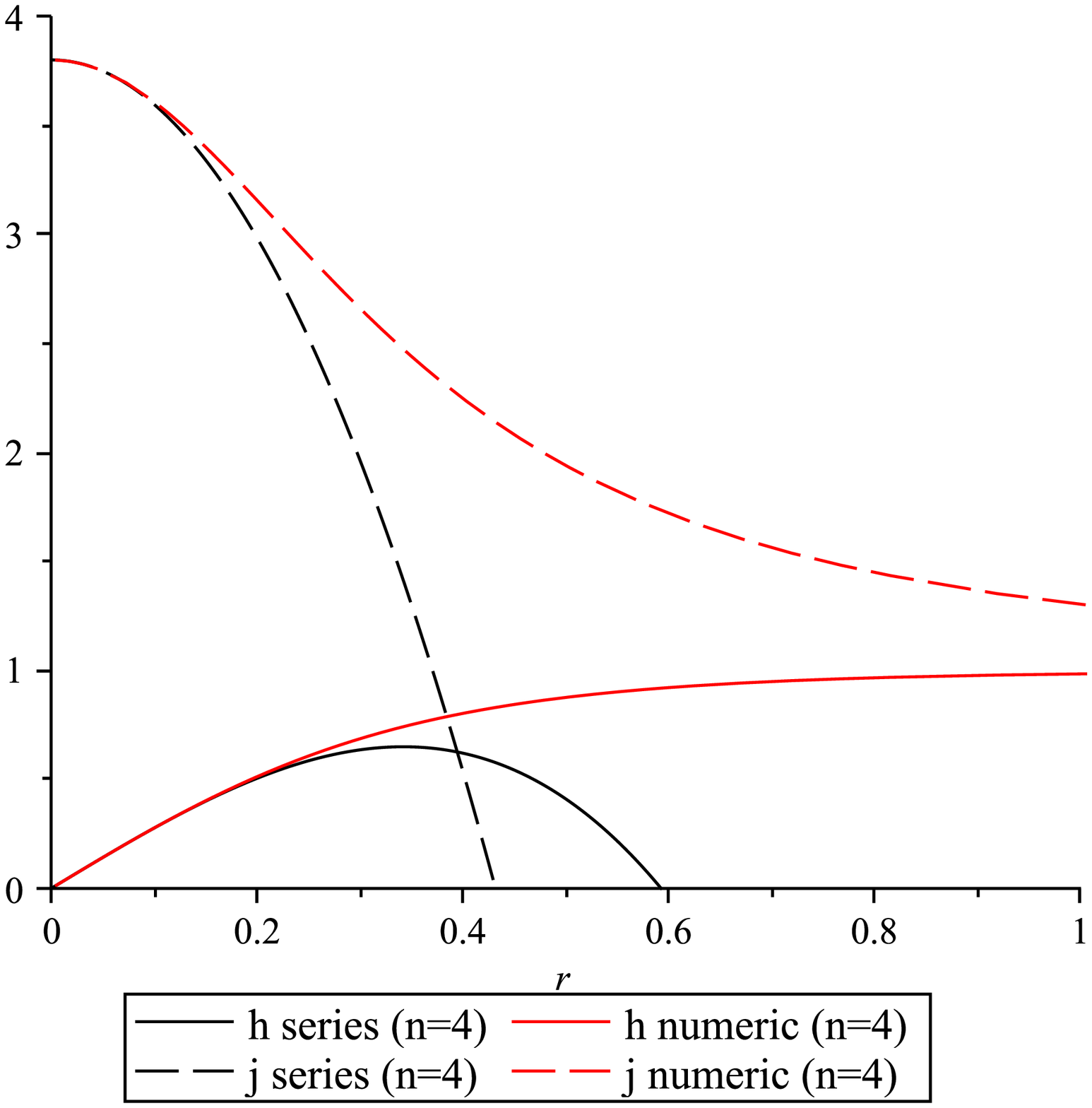}
\par\end{center}

\noindent \textbf{Figure 7:} The metric and gauge functions corresponding
to the lowest magic value for $n=4$: numeric solutions (Figure 5a)
versus series solutions ((\ref{eq:18})-(\ref{eq:21}), truncated
after two terms) for small $r$. Left: The metric functions $f\left(r\right)$
and $g\left(r\right)$. Right: The gauge functions $h\left(r\right)$
and $j\left(r\right)$. 

\bibliography{references}
\bibliographystyle{unsrt}

\end{document}